\documentclass[twocolumn,showpacs,babel,superscriptaddress,german,final,prl]{revtex4}
\usepackage{graphicx}
\usepackage{amsmath}

\begin{document}

\author{Thomas Juffmann}%

\author{Stefan Truppe}%

\author{Philipp Geyer}%
\author{Andr\'as G. Major}%
\altaffiliation{Current address: Carl Zeiss  SMC AG,
Rudolf-Eber-Stra{\ss}e 2, 73447 Oberkochen, Germany}
\affiliation{Faculty of Physics, University of Vienna,
Boltzmanngasse 5, 1090 Vienna, Austria}

\author{Sarayut Deachapunya}%
\affiliation{Faculty of Physics, University of Vienna,
Boltzmanngasse 5, 1090 Vienna, Austria}
\affiliation{Department of Physics, Faculty of Science, Burapha University,
 Chonburi 20131, Thailand}

\author{Hendrik Ulbricht}%
\affiliation{Faculty of Physics, University of Vienna,
Boltzmanngasse 5, 1090 Vienna, Austria}
\affiliation{School of Physics and Astronomy, University of Southampton, SO17 1BJ, UK}

\author{Markus Arndt}%
\affiliation{Faculty of Physics, University of Vienna,
Boltzmanngasse 5, 1090 Vienna, Austria}
\homepage{http://www.quantum.at} \email{markus.arndt@univie.ac.at}

\title{Wave and particle in molecular interference lithography}

\date{\today}
\begin{abstract}
The wave-particle duality of massive objects is a cornerstone of quantum physics and a key property of many modern tools such as electron microscopy, neutron diffraction or atom interferometry.
Here we report on the first experimental demonstration of quantum interference lithography with complex molecules. Molecular matter-wave interference patterns are deposited onto a reconstructed Si(111)\,7$\times$7 surface and imaged using scanning tunneling microscopy. Thereby both the particle and the quantum wave character of the molecules can be visualized in one and the same image. This new approach to nanolithography therefore also represents a sensitive new detection scheme for quantum interference experiments.
\end{abstract}


\maketitle

The de Broglie wave nature of massive particles has always been an essential ingredient in the conceptual development of quantum mechanics~\cite{DeBroglie1923a,Schroedinger1926a}.
First demonstrations of the electron wave nature~\cite{Davisson1927a,Thomson1927a} were soon followed by experiments on the diffraction of helium atoms and H$_{2}$ molecules~\cite{Estermann1930a} as well as neutrons~\cite{Halban1936,Rauch2000a}.
The build-up of quantum interference patterns from single particles was shown in particularly nice demonstrations with individual photons~\cite{Taylor1909a} and electrons~\cite{Tonomura1989a,Hasselbach1997a}.
With the  availability of laser and nanofabrication technologies, atom interferometry and coherent lithography have become rapidly developing fields of research~\cite{Berman1997a,Shimizu2002a,Oberthaler2003a,Cronin2009a}.
Recently, quantum interference experiments have also been extended to composite nanoparticles, such as fullerenes~\cite{Arndt1999a}, He clusters~\cite{Bruehl2004a} or large fluorinated molecules~\cite{Hackermuller2003a,Gerlich2008a}.

Our present demonstration complements these earlier studies as it represents the first realization of quantum interference lithography with molecules. It thereby closes two gaps that so far existed in the field of macromolecule interferometry: On the one hand, previous experiments had not been able to visualize the individual particles that traversed the interferometer. Further, it has also been discussed by several authors that ionizing detectors are inefficient already for medium-sized organic materials ~\cite{Marksteiner2008a,Hanley2009a}.
Being sensitive to single molecules, our new lithographic interference detection scheme provides, for the first time, the opportunity to visualize both the quantum wave features and the composite particle nature of individual molecules in one and the same image and experiment.
On the other hand, we show that near-field interferometry is  a very natural approach to generating surface-deposited and immobilized nanopatterns with particles that can, in principle, be  regarded as functional entities by themselves. Our experiment  thus operates at the interface between matter-wave interferometry and surface nanoscience.

A schematic illustration of the setup is shown in Figure 1. The experiment
is sectioned into five differentially pumped vacuum chambers, which comprise three logical compartments: the source, the interferometric nanodeposition and the detector. The supplementary vacuum chambers are required for differential pumping as well as for the preparation and transfer of the detection surface.
A Knudsen cell at T=1070\,K forms the molecular beam which is  selected within  a transmitted velocity band of $\Delta v/v= 0.05$\,(FWHM) to account for the fact that different velocities correspond to different de Broglie wavelengths, $\lambda_{\mathrm{dB}}=h/mv$, and to varying interaction times with the gratings.

Two alternative methods have been used for selecting the molecular speed:
First, a gravitational velocity selection scheme exploits the molecular free-fall trajectories in the Earth's gravitational field. This established method~\cite{Nairz2001a,Brezger2002a} allows to work without any vibrating or rotating element. The interference fringe contrast of the molecular deposit varies, however, with the vertical position on the sample and scattering at the slit edges may spoil the selection quality.

Second, interferograms were also recorded using a home-built helical velocity selector which is a miniaturized version of devices known from neutron scattering~\cite{Miller1955a}. Grooves of 300\,$\mu$m width were milled along helical trajectories into an aluminum cup of 40\,mm  length. The aspect ratio of length and width of the grooves defines the transmitted velocity bandwidth. Their pitch and angular speed defines the center of the distribution. The selector is held by UHV compatible bearings and driven by a motor outside of the chamber. A magneto-fluidic seal allows the mechanical transduction of the rotational motion at frequencies  in excess of 100\,Hz.  An acceleration sensor at the mechanical base of the interferometer quantified typical oscillation amplitudes to be as small as 2\,nm.

This method was chosen for the experiment presented here. Its advantage lies in the fact that all transmitted particles arrive with the same speed and the nanostructure has the same contrast across the entire surface that is illuminated by the molecular beam. Being independent of gravitation the selection scheme can be used in any  orientation of the experiment.

About 110\,cm behind the source, the molecules encounter the first diffraction grating $G_{1}$ of a near-field interferometer~\cite{Clauser1994a,Brezger2002a,Nimmrichter2008a}. Diffraction at each of the individual slits within $G_{1}$ expands the
molecular coherence function to an extent that it covers several slits of the second grating $G_{2}$.
 Diffraction at $G_{2}$, coherent evolution and interference subsequently generate a molecular density pattern, which we accumulate on the detector screen $D$.

The SiN$_{x}$ gratings were  fabricated by Dr.\,Savas at MIT, Cambridge~\cite{Gerlich2007a} with a highly accurate period of $d=257.40(1)$\,nm and with open slit windows as small as 75\,nm in G$_{1}$ and 150\,nm in G$_{2}$.
In our symmetrical setup, the separation $L$ between the two gratings is equal to the distance between G$_{2}$ and $D$.
The molecular distribution on the detector is then an approximate  self-image of the transmission function of $G_{2}$, if the Talbot condition is met, i.e. if  $L= d^2/\lambda_{\mathrm{dB}}$ in the absence of any external potentials.
Although the nanomechanical gratings are fabricated with a rectangular transmission profile, interference and phase shifts in the gratings lead to a near-sinusoidal molecular density pattern on the screen. The fringe visibility can then be extracted as $V=(S_{\mathrm{max}}-S_{\mathrm{min}})/(S_{\mathrm{max}}+S_{\mathrm{min}})$, where $S(x)$ is the local molecular surface density at position $x$.

The detector screen is a thermally reconstructed Si\,(111)\,7$\times7$ surface. Silicon is known to capture and bind fullerenes exceptionally well~\cite{Chen1994a} and it is also the natural choice for interfacing molecular deposits to future electronic readout or control. The atomically clean and flat surface immobilizes the incident molecules and it allows us to operate the experiment at room temperature. It requires, however, a proper surface cleaning, preheating and flash-heating in the read-out chamber which also contains the variable temperature scanning tunneling microscope(RHK STM UHV\,700).

For a grating separation of $L=13.2$\,mm we expect the maximum interference contrast of the first Talbot order for a de Broglie wavelength of 5\,pm, i.e. for C$_{60}$ with a mean velocity of 111\,m/s.  The van der Waals interaction between the molecule and the grating wall reduces the effective grating opening and shifts the theoretical fringe visibility. At a mean molecular velocity of $v=115$\,m/s we expect a maximum contrast of about $V=60\,\%$.
If the molecules were classical billiard balls, i.e. following trajectories of Newtonian physics, theory would predict a nearly flat molecular distribution at the detector, i.e. a vanishing fringe visibility of only $V=1\,\%$.
This comparison of visibilities underlines the importance of the molecular quantum wave nature for the emergence of high-contrast patterns in this kind of lithography.

\begin{figure}
  \includegraphics[width=1\columnwidth]{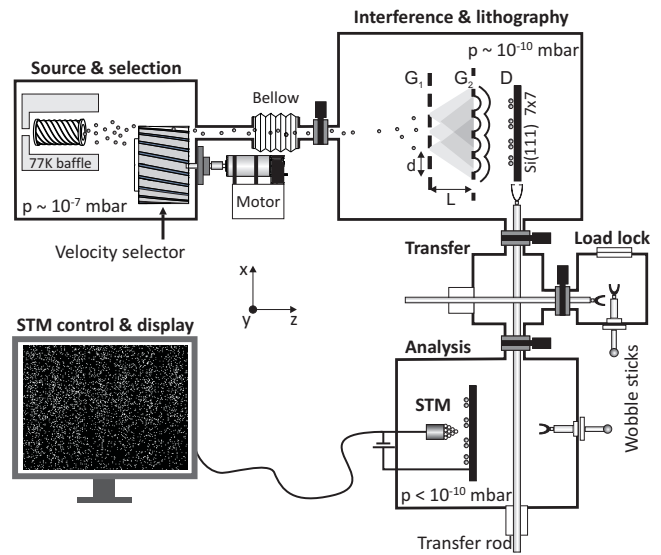}
  \caption{Molecular interference lithography combines coherent molecule wave propagation with surface deposition and scanning probe microscopy.
  The furnace emits a thermal fullerene beam, velocity selected to
  $\Delta v/v = 5\%$, which passes two SiN gratings ($d=257.40$\,nm) separated by $L=12.5$\,mm. A prepared silicon sample is placed in distance $L$ behind $G_{2}$. The molecular deposit is imaged with single molecule resolution using scanning tunneling microscopy in a separate UHV chamber. }\label{Fig1}
\end{figure}

The total experimental sequence is as follows:  A silicon surface is thermally reconstructed, characterized in the STM and then transferred into the interferometer. The angular frequency of the helical velocity selector is set to transmit the velocity class $v=115\pm5$\,m/s. The molecular exposure lasts about 30\,min and we deposit about 0.001 molecular monolayers (ML) on the target. This low coverage is required to maintain  the single-particle character in the quantum demonstration. The exposed sample is then transferred into the detector chamber.

The STM surface scan of  Figure\,2 clearly reveals the individual silicon substrate atoms as well as several immobilized single C$_{60}$ molecules.
The surface binding of the fullerenes is so strong that we could not observe any clustering, even over two weeks. High-resolution tunneling microscopy at low temperatures even allows us to get a glimpse on the {\em internal} structure of the deposited fullerenes~\cite{Hou1999a}, as shown in the inset of Figure\,2.

\begin{figure}
\centering
  \includegraphics[width=0.8\columnwidth]{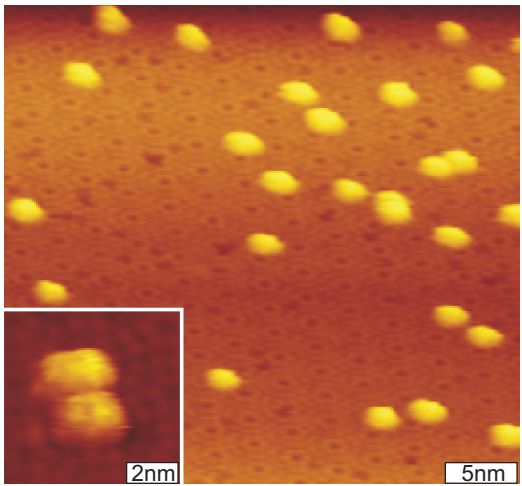}
  \caption{STM image of a reconstructed Si (111) $7\times7$ surface, covered with immobilized individual fullerenes. Area of this image: $33\times36$\,nm$^2$; tunneling current I=0.2\,nA; sample bias voltage U=+2V; temperature T=30\,K.  A close up on two fullerenes (see inset) gives a glimpse on the internal ring structure of C$_{60}$ (see also~\cite{Hou1999a}).}\label{Fig2}
\end{figure}

In order to see the interference pattern we probe an area of 2\,$\mu$m$^2$ at a resolution that still allows to identify the individual molecules. This scan takes about thirty minutes and returns 10$^6$ data points.
We identify the molecules by their height: if a recorded pixel (i,j) is 0.5-0.9\,nm higher than the average of the neighboring pixels ($i\pm 3,j\pm 3$) it is identified as a fullerene molecule.
The analog images are thus converted into a binary matrix, where the presence of a molecule is represented by the digit one, whereas all other pixels are set to zero. The result of this procedure is shown in Figure\,3a, where the bright dots represent the molecules. For a better visualization, the pixels were later binned in $5\times5$ boxes. The resulting image reveals five interference fringes in the chosen section.

In order to quantify the fringe contrast we perform a vertical sum (along the y-direction) over all rows and arrive at the one-dimensional interference curve. Since the statistical fluctuations are still rather high, with only 1166 molecules per \,$\mu$m$^2$, we sum over twenty horizontal points (along the x-direction) to obtain Figure\,3b. The data are well represented by a numerical fit of the form $F(x)=A\sin(2\pi x/d+\phi_0)+B$, from which we extract a fringe visibility of $V=A/B=36\pm3\,\%$. The statistical error is computed from a Levenberg-Marquardt algorithm which is weighted with the Poissonian uncertainty of each individual column in the picture.
A similar interference contrast was found in the gravitational velocity selection scheme.  This value is lower than the theoretical expectation, but already a linear drift of 2.5\,nm per minute of either one grating or the surface is sufficient to explain this observation. In future experiments, thermal drifts can be further reduced by replacing the interferometer support structure by materials of lower thermal expansion.

The recorded lattice has a period of 267(3)\,nm which differs slightly from the  expected 257\,nm. The 4\,\% period mismatch is consistent with a linear thermal drift rate, now in the STM instead of the interferometer, as small as 0.4\,nm per minute.  The experiment was repeated several times, yielding an interference pattern equally close to the expected period and orientation.

The observed fringe contrast exceeds the classical moir\'{e} expectation in the presence of van der Waals forces by more than a factor of thirty, and the experiment is offset from the classical result by ten standard deviations. Quantum delocalization and interference are therefore needed for creating high contrast molecular nanopatterns using non-contact mask imaging.
Surface adsorption and imaging is also an intuitive tool in experiments on the foundations of physics, as it allows to visualize the localized but random particle positions within a deterministic fringe pattern that is prescribed by the free evolution of the delocalized matter wave (Figure 3).
The surface probe technique is capable of detecting individual molecules and it is well-suited for future experiments with velocity selected, monodisperse beams of much larger objects in different interferometer configurations~\cite{Gerlich2007a}.

\begin{figure}
\centering
  \includegraphics[width=0.9\columnwidth]{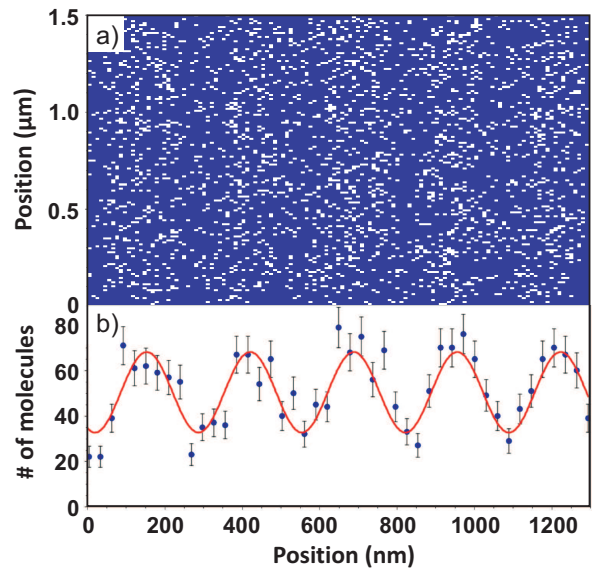}\\
  \caption{a) The processed STM image (see text) reveals both the particle-nature and the quantum wave-nature of the surface deposited fullerene molecules in one and the same image. b) A vertical sum over (a) yields the interference curve. The fringe visibility amounts to 36\%. The error bars represent the $\sqrt{N}$-noise related to the small number $N$ of particles per bin. }\label{Fig3}
\end{figure}

Single molecules may be regarded as functional elements. They may serve as immobilized single-photon emitters~\cite{Lounis2005a}, organic switches~\cite{Joachim2000a}, nanomachines~\cite{Koumura1999a,Browne2006a}, transistor components~\cite{Park2000a} or as nucleation cores for the catalysis of molecular growth. Positioning them on surfaces might thus be crucial in future applications in nanotechnology.
Our molecule lithography scheme works at a de Broglie wavelength comparable to that prevailing in high-energy electron beam writing. The kinetic energy E$_\mathrm{kin}$=0.1\,eV and the velocity $v=100$\,m/s can, however, be many orders of magnitude smaller than that of the electrons.
It therefore combines the potential of high resolution with minimal damage to the surface.

Future experiments shall explore how to build more complex molecular patterns in this non-contact, constructive and parallel  way.
High-contrast interference provides the possibility to exclude molecules with certainty from some surface areas. The positioning within an interference maximum is, however, still affected by the probabilistic character of the quantum wavefunction. The combination of interferometric prestructuring with local self-organization or STM postprocessing~\cite{Beton1995a} appears to be a way towards deterministic molecular nanostructures~\cite{Barth2005a}.
Our present proof-of-principle demonstration shows the interferometric generation of molecular lines. Using two-dimensional cross gratings and either electric deflectometry~\cite{Berninger2007a} or motorized gratings it will be possible to write periodic arrays of more complex patterns, too~\cite{Patorski1989a}.
In an asymmetric Talbot-Lau interferometer the exploitation of the fractional Talbot effect shall further allow writing of structures smaller than the grating period~\cite{Brezger2003a}, eventually smaller than all features in the lithography masks.

This work has been supported by the FWF within the
Wittgenstein project Z149-N16. A.M. acknowledges support
by the FWF Lise-Meitner program M887-N02. We
acknowledge support by the ESF EUROCORE Program
EUROQUASAR MIME.


\begin{thebibliography}{38}
\expandafter\ifx\csname natexlab\endcsname\relax\def\natexlab#1{#1}\fi
\expandafter\ifx\csname bibnamefont\endcsname\relax
  \def\bibnamefont#1{#1}\fi
\expandafter\ifx\csname bibfnamefont\endcsname\relax
  \def\bibfnamefont#1{#1}\fi
\expandafter\ifx\csname citenamefont\endcsname\relax
  \def\citenamefont#1{#1}\fi
\expandafter\ifx\csname url\endcsname\relax
  \def\url#1{\texttt{#1}}\fi
\expandafter\ifx\csname urlprefix\endcsname\relax\def\urlprefix{URL }\fi
\providecommand{\bibinfo}[2]{#2}
\providecommand{\eprint}[2][]{\url{#2}}

\bibitem[{\citenamefont{de~Broglie}(1923)}]{DeBroglie1923a}
\bibinfo{author}{\bibfnamefont{L.}~\bibnamefont{de~Broglie}},
  \bibinfo{journal}{Nature} \textbf{\bibinfo{volume}{112}},
  \bibinfo{pages}{540} (\bibinfo{year}{1923}).

\bibitem[{\citenamefont{Schr\"{o}dinger}(1926)}]{Schroedinger1926a}
\bibinfo{author}{\bibfnamefont{E.}~\bibnamefont{Schr\"{o}dinger}},
  \bibinfo{journal}{Annalen der Physik} \textbf{\bibinfo{volume}{79}},
  \bibinfo{pages}{361} (\bibinfo{year}{1926}).

\bibitem[{\citenamefont{Davisson and Germer}(1927)}]{Davisson1927a}
\bibinfo{author}{\bibfnamefont{C.}~\bibnamefont{Davisson}} \bibnamefont{and}
  \bibinfo{author}{\bibfnamefont{L.}~\bibnamefont{Germer}},
  \bibinfo{journal}{Nature} \textbf{\bibinfo{volume}{119}}, \bibinfo{pages}{558
  } (\bibinfo{year}{1927}).

\bibitem[{\citenamefont{Thomson}(1927)}]{Thomson1927a}
\bibinfo{author}{\bibfnamefont{G.~P.} \bibnamefont{Thomson}},
  \bibinfo{journal}{Nature} \textbf{\bibinfo{volume}{120}},
  \bibinfo{pages}{802} (\bibinfo{year}{1927}).

\bibitem[{\citenamefont{Estermann and Stern}(1930)}]{Estermann1930a}
\bibinfo{author}{\bibfnamefont{I.}~\bibnamefont{Estermann}} \bibnamefont{and}
  \bibinfo{author}{\bibfnamefont{O.}~\bibnamefont{Stern}}, \bibinfo{journal}{Z.
  Phys.} \textbf{\bibinfo{volume}{61}}, \bibinfo{pages}{95}
  (\bibinfo{year}{1930}).

\bibitem[{\citenamefont{von Halban and Preiswerk}(1936)}]{Halban1936}
\bibinfo{author}{\bibfnamefont{H.}~\bibnamefont{von Halban}} \bibnamefont{and}
  \bibinfo{author}{\bibfnamefont{P.}~\bibnamefont{Preiswerk}},
  \bibinfo{journal}{C.\ R.\ Hebd.\ S\'{e}ances Acad.}
  \textbf{\bibinfo{volume}{203}}, \bibinfo{pages}{73} (\bibinfo{year}{1936}).

\bibitem[{\citenamefont{Rauch and Werner}(2000)}]{Rauch2000a}
\bibinfo{author}{\bibfnamefont{H.}~\bibnamefont{Rauch}} \bibnamefont{and}
  \bibinfo{author}{\bibfnamefont{A.}~\bibnamefont{Werner}},
  \emph{\bibinfo{title}{Neutron Interferometry: Lessons in Experimental Quantum
  Mechanics}} (\bibinfo{publisher}{Oxford Univ. Press}, \bibinfo{year}{2000}).

\bibitem[{\citenamefont{I.Taylor}(1909)}]{Taylor1909a}
\bibinfo{author}{\bibfnamefont{G.}~\bibnamefont{I.Taylor}},
  \bibinfo{journal}{Proc. Cambridge Phil. Soc.} \textbf{\bibinfo{volume}{15}},
  \bibinfo{pages}{114} (\bibinfo{year}{1909}).

\bibitem[{\citenamefont{{Tonomura} et~al.}(1989)\citenamefont{{Tonomura},
  {Endo}, {Matsuda}, {Kawasaki}, and {Ezawa}}}]{Tonomura1989a}
\bibinfo{author}{\bibfnamefont{A.}~\bibnamefont{{Tonomura}}},
  \bibinfo{author}{\bibfnamefont{J.}~\bibnamefont{{Endo}}},
  \bibinfo{author}{\bibfnamefont{T.}~\bibnamefont{{Matsuda}}},
  \bibinfo{author}{\bibfnamefont{T.}~\bibnamefont{{Kawasaki}}},
  \bibnamefont{and} \bibinfo{author}{\bibfnamefont{H.}~\bibnamefont{{Ezawa}}},
  \bibinfo{journal}{Am. J. Phys.} \textbf{\bibinfo{volume}{57}},
  \bibinfo{pages}{117} (\bibinfo{year}{1989}).

\bibitem[{\citenamefont{Hasselbach}(1997)}]{Hasselbach1997a}
\bibinfo{author}{\bibfnamefont{F.}~\bibnamefont{Hasselbach}},
  \bibinfo{journal}{Scanning Microscopy} \textbf{\bibinfo{volume}{11}},
  \bibinfo{pages}{345 } (\bibinfo{year}{1997}).

\bibitem[{\citenamefont{Berman}(1997)}]{Berman1997a}
\bibinfo{author}{\bibfnamefont{P.~R.} \bibnamefont{Berman}},
  \emph{\bibinfo{title}{Atom Interferometry}} (\bibinfo{publisher}{Acad.
  Press}, \bibinfo{address}{New York}, \bibinfo{year}{1997}).

\bibitem[{\citenamefont{Shimizu and Fujita}(2002)}]{Shimizu2002a}
\bibinfo{author}{\bibfnamefont{F.}~\bibnamefont{Shimizu}} \bibnamefont{and}
  \bibinfo{author}{\bibfnamefont{J.}~\bibnamefont{Fujita}},
  \bibinfo{journal}{Phys. Rev. Lett} \textbf{\bibinfo{volume}{88}},
  \bibinfo{pages}{123201} (\bibinfo{year}{2002}).

\bibitem[{\citenamefont{Oberthaler and Pfau}(2003)}]{Oberthaler2003a}
\bibinfo{author}{\bibfnamefont{M.}~\bibnamefont{Oberthaler}} \bibnamefont{and}
  \bibinfo{author}{\bibfnamefont{T.}~\bibnamefont{Pfau}}, \bibinfo{journal}{J.
  Phys. Condensed Matter} \textbf{\bibinfo{volume}{15}}, \bibinfo{pages}{R233}
  (\bibinfo{year}{2003}).

\bibitem[{\citenamefont{Cronin et~al.}(2009)\citenamefont{Cronin, Schmiedmayer,
  and Pritchard}}]{Cronin2009a}
\bibinfo{author}{\bibfnamefont{A.~D.} \bibnamefont{Cronin}},
  \bibinfo{author}{\bibfnamefont{J.}~\bibnamefont{Schmiedmayer}},
  \bibnamefont{and} \bibinfo{author}{\bibfnamefont{D.~E.}
  \bibnamefont{Pritchard}}, \bibinfo{journal}{Rev. Mod. Phys.}
  \textbf{\bibinfo{volume}{81}}, \bibinfo{pages}{1051} (\bibinfo{year}{2009}).

\bibitem[{\citenamefont{Arndt et~al.}(1999)\citenamefont{Arndt, Nairz,
  Voss-Andreae, Keller, der Zouw, and Zeilinger}}]{Arndt1999a}
\bibinfo{author}{\bibfnamefont{M.}~\bibnamefont{Arndt}},
  \bibinfo{author}{\bibfnamefont{O.}~\bibnamefont{Nairz}},
  \bibinfo{author}{\bibfnamefont{J.}~\bibnamefont{Voss-Andreae}},
  \bibinfo{author}{\bibfnamefont{C.}~\bibnamefont{Keller}},
  \bibinfo{author}{\bibfnamefont{G.~V.} \bibnamefont{der Zouw}},
  \bibnamefont{and}
  \bibinfo{author}{\bibfnamefont{A.}~\bibnamefont{Zeilinger}},
  \bibinfo{journal}{Nature} \textbf{\bibinfo{volume}{401}},
  \bibinfo{pages}{680} (\bibinfo{year}{1999}).

\bibitem[{\citenamefont{Bruehl et~al.}(2004)\citenamefont{Bruehl, Guardiola,
  Kalinin, Kornilov, Navarro, Savas, and Toennies}}]{Bruehl2004a}
\bibinfo{author}{\bibfnamefont{R.}~\bibnamefont{Bruehl}},
  \bibinfo{author}{\bibfnamefont{R.}~\bibnamefont{Guardiola}},
  \bibinfo{author}{\bibfnamefont{A.}~\bibnamefont{Kalinin}},
  \bibinfo{author}{\bibfnamefont{O.}~\bibnamefont{Kornilov}},
  \bibinfo{author}{\bibfnamefont{J.}~\bibnamefont{Navarro}},
  \bibinfo{author}{\bibfnamefont{T.}~\bibnamefont{Savas}}, \bibnamefont{and}
  \bibinfo{author}{\bibfnamefont{J.~P.} \bibnamefont{Toennies}},
  \bibinfo{journal}{Phys. Rev. Lett.} \textbf{\bibinfo{volume}{92}},
  \bibinfo{pages}{185301} (\bibinfo{year}{2004}).

\bibitem[{\citenamefont{Hackerm\"{u}ller
  et~al.}(2003)\citenamefont{Hackerm\"{u}ller, Uttenthaler, Hornberger, Reiger,
  Brezger, Zeilinger, and Arndt}}]{Hackermuller2003a}
\bibinfo{author}{\bibfnamefont{L.}~\bibnamefont{Hackerm\"{u}ller}},
  \bibinfo{author}{\bibfnamefont{S.}~\bibnamefont{Uttenthaler}},
  \bibinfo{author}{\bibfnamefont{K.}~\bibnamefont{Hornberger}},
  \bibinfo{author}{\bibfnamefont{E.}~\bibnamefont{Reiger}},
  \bibinfo{author}{\bibfnamefont{B.}~\bibnamefont{Brezger}},
  \bibinfo{author}{\bibfnamefont{A.}~\bibnamefont{Zeilinger}},
  \bibnamefont{and} \bibinfo{author}{\bibfnamefont{M.}~\bibnamefont{Arndt}},
  \bibinfo{journal}{Phys. Rev. Lett.} \textbf{\bibinfo{volume}{91}}
  (\bibinfo{year}{2003}).

\bibitem[{\citenamefont{Gerlich et~al.}(2008)\citenamefont{Gerlich, Gring,
  Ulbricht, Hornberger, T\"{u}xen, Mayor, and Arndt}}]{Gerlich2008a}
\bibinfo{author}{\bibfnamefont{S.}~\bibnamefont{Gerlich}},
  \bibinfo{author}{\bibfnamefont{M.}~\bibnamefont{Gring}},
  \bibinfo{author}{\bibfnamefont{H.}~\bibnamefont{Ulbricht}},
  \bibinfo{author}{\bibfnamefont{K.}~\bibnamefont{Hornberger}},
  \bibinfo{author}{\bibfnamefont{J.}~\bibnamefont{T\"{u}xen}},
  \bibinfo{author}{\bibfnamefont{M.}~\bibnamefont{Mayor}}, \bibnamefont{and}
  \bibinfo{author}{\bibfnamefont{M.}~\bibnamefont{Arndt}},
  \bibinfo{journal}{Angew. Chem. Int. Ed.} \textbf{\bibinfo{volume}{47}},
  \bibinfo{pages}{6195 } (\bibinfo{year}{2008}).

\bibitem[{\citenamefont{Marksteiner et~al.}(2008)\citenamefont{Marksteiner,
  Haslinger, Ulbricht, Sclafani, Oberhofer, Dellago, and
  Arndt}}]{Marksteiner2008a}
\bibinfo{author}{\bibfnamefont{M.}~\bibnamefont{Marksteiner}},
  \bibinfo{author}{\bibfnamefont{P.}~\bibnamefont{Haslinger}},
  \bibinfo{author}{\bibfnamefont{H.}~\bibnamefont{Ulbricht}},
  \bibinfo{author}{\bibfnamefont{M.}~\bibnamefont{Sclafani}},
  \bibinfo{author}{\bibfnamefont{H.}~\bibnamefont{Oberhofer}},
  \bibinfo{author}{\bibfnamefont{C.}~\bibnamefont{Dellago}}, \bibnamefont{and}
  \bibinfo{author}{\bibfnamefont{M.}~\bibnamefont{Arndt}}, \bibinfo{journal}{J.
  Am. Soc. Mass. Spectrom.} \textbf{\bibinfo{volume}{19}}, \bibinfo{pages}{1021
  } (\bibinfo{year}{2008}).

\bibitem[{\citenamefont{Hanley and Zimmermann}(2009)}]{Hanley2009a}
\bibinfo{author}{\bibfnamefont{L.}~\bibnamefont{Hanley}} \bibnamefont{and}
  \bibinfo{author}{\bibfnamefont{R.}~\bibnamefont{Zimmermann}},
  \bibinfo{journal}{Analytical Chemistry} \textbf{\bibinfo{volume}{81}},
  \bibinfo{pages}{4174} (\bibinfo{year}{2009}).

\bibitem[{\citenamefont{Nairz et~al.}(2001)\citenamefont{Nairz, Brezger, Arndt,
  and Zeilinger}}]{Nairz2001a}
\bibinfo{author}{\bibfnamefont{O.}~\bibnamefont{Nairz}},
  \bibinfo{author}{\bibfnamefont{B.}~\bibnamefont{Brezger}},
  \bibinfo{author}{\bibfnamefont{M.}~\bibnamefont{Arndt}}, \bibnamefont{and}
  \bibinfo{author}{\bibfnamefont{A.}~\bibnamefont{Zeilinger}},
  \bibinfo{journal}{Phys. Rev. Lett.} \textbf{\bibinfo{volume}{87}},
  \bibinfo{pages}{160401} (\bibinfo{year}{2001}).

\bibitem[{\citenamefont{Brezger et~al.}(2002)\citenamefont{Brezger,
  Hackerm\"{u}ller, Uttenthaler, Petschinka, Arndt, and
  Zeilinger}}]{Brezger2002a}
\bibinfo{author}{\bibfnamefont{B.}~\bibnamefont{Brezger}},
  \bibinfo{author}{\bibfnamefont{L.}~\bibnamefont{Hackerm\"{u}ller}},
  \bibinfo{author}{\bibfnamefont{S.}~\bibnamefont{Uttenthaler}},
  \bibinfo{author}{\bibfnamefont{J.}~\bibnamefont{Petschinka}},
  \bibinfo{author}{\bibfnamefont{M.}~\bibnamefont{Arndt}}, \bibnamefont{and}
  \bibinfo{author}{\bibfnamefont{A.}~\bibnamefont{Zeilinger}},
  \bibinfo{journal}{Phys. Rev. Lett.} \textbf{\bibinfo{volume}{88}},
  \bibinfo{pages}{100404} (\bibinfo{year}{2002}).

\bibitem[{\citenamefont{Miller and Kusch}(1955)}]{Miller1955a}
\bibinfo{author}{\bibfnamefont{R.}~\bibnamefont{Miller}} \bibnamefont{and}
  \bibinfo{author}{\bibfnamefont{P.}~\bibnamefont{Kusch}},
  \bibinfo{journal}{Phys. Rev.} \textbf{\bibinfo{volume}{99}},
  \bibinfo{pages}{1314} (\bibinfo{year}{1955}).

\bibitem[{\citenamefont{Clauser and Li}(1994)}]{Clauser1994a}
\bibinfo{author}{\bibfnamefont{J.~F.} \bibnamefont{Clauser}} \bibnamefont{and}
  \bibinfo{author}{\bibfnamefont{S.}~\bibnamefont{Li}},
  \bibinfo{journal}{Phys.\ Rev.\ A} \textbf{\bibinfo{volume}{49}},
  \bibinfo{pages}{R2213} (\bibinfo{year}{1994}).

\bibitem[{\citenamefont{Nimmrichter and Hornberger}(2008)}]{Nimmrichter2008a}
\bibinfo{author}{\bibfnamefont{S.}~\bibnamefont{Nimmrichter}} \bibnamefont{and}
  \bibinfo{author}{\bibfnamefont{K.}~\bibnamefont{Hornberger}},
  \bibinfo{journal}{Phys. Rev. A} \textbf{\bibinfo{volume}{78}},
  \bibinfo{pages}{023612} (\bibinfo{year}{2008}).

\bibitem[{\citenamefont{Gerlich et~al.}(2007)\citenamefont{Gerlich,
  Hackerm\"uller, Hornberger, Stibor, Ulbricht, Gring, Goldfarb, Savas,
  M\"{u}ri, Mayor et~al.}}]{Gerlich2007a}
\bibinfo{author}{\bibfnamefont{S.}~\bibnamefont{Gerlich}},
  \bibinfo{author}{\bibfnamefont{L.}~\bibnamefont{Hackerm\"uller}},
  \bibinfo{author}{\bibfnamefont{K.}~\bibnamefont{Hornberger}},
  \bibinfo{author}{\bibfnamefont{A.}~\bibnamefont{Stibor}},
  \bibinfo{author}{\bibfnamefont{H.}~\bibnamefont{Ulbricht}},
  \bibinfo{author}{\bibfnamefont{M.}~\bibnamefont{Gring}},
  \bibinfo{author}{\bibfnamefont{F.}~\bibnamefont{Goldfarb}},
  \bibinfo{author}{\bibfnamefont{T.}~\bibnamefont{Savas}},
  \bibinfo{author}{\bibfnamefont{M.}~\bibnamefont{M\"{u}ri}},
  \bibinfo{author}{\bibfnamefont{M.}~\bibnamefont{Mayor}},
  \bibnamefont{et~al.}, \bibinfo{journal}{Nature Phys.}
  \textbf{\bibinfo{volume}{3}}, \bibinfo{pages}{711 } (\bibinfo{year}{2007}).

\bibitem[{\citenamefont{Chen and Sarid}(1994)}]{Chen1994a}
\bibinfo{author}{\bibfnamefont{D.}~\bibnamefont{Chen}} \bibnamefont{and}
  \bibinfo{author}{\bibfnamefont{D.}~\bibnamefont{Sarid}},
  \bibinfo{journal}{Surf. Sci.} \textbf{\bibinfo{volume}{318}},
  \bibinfo{pages}{74} (\bibinfo{year}{1994}).

\bibitem[{\citenamefont{Hou et~al.}(1999)\citenamefont{Hou, Jinlong, Haiqian,
  Qunxiang, Changgan, Hai, Wang, Chen, and Qingshi}}]{Hou1999a}
\bibinfo{author}{\bibfnamefont{J.~G.} \bibnamefont{Hou}},
  \bibinfo{author}{\bibfnamefont{Y.}~\bibnamefont{Jinlong}},
  \bibinfo{author}{\bibfnamefont{W.}~\bibnamefont{Haiqian}},
  \bibinfo{author}{\bibfnamefont{L.}~\bibnamefont{Qunxiang}},
  \bibinfo{author}{\bibfnamefont{Z.}~\bibnamefont{Changgan}},
  \bibinfo{author}{\bibfnamefont{L.}~\bibnamefont{Hai}},
  \bibinfo{author}{\bibfnamefont{B.}~\bibnamefont{Wang}},
  \bibinfo{author}{\bibfnamefont{D.~M.} \bibnamefont{Chen}}, \bibnamefont{and}
  \bibinfo{author}{\bibfnamefont{Z.}~\bibnamefont{Qingshi}},
  \bibinfo{journal}{Phys. Rev. Lett.} \textbf{\bibinfo{volume}{83}},
  \bibinfo{pages}{3001} (\bibinfo{year}{1999}).

\bibitem[{\citenamefont{Lounis and Orrit}(2005)}]{Lounis2005a}
\bibinfo{author}{\bibfnamefont{B.}~\bibnamefont{Lounis}} \bibnamefont{and}
  \bibinfo{author}{\bibfnamefont{M.}~\bibnamefont{Orrit}},
  \bibinfo{journal}{Rep. Prog. Phys.} \textbf{\bibinfo{volume}{68}},
  \bibinfo{pages}{1129–} (\bibinfo{year}{2005}).

\bibitem[{\citenamefont{Joachim et~al.}(2000)\citenamefont{Joachim, \&, and
  Aviram}}]{Joachim2000a}
\bibinfo{author}{\bibfnamefont{C.}~\bibnamefont{Joachim}},
  \bibinfo{author}{\bibfnamefont{J.~K.~G.} \bibnamefont{\&}}, \bibnamefont{and}
  \bibinfo{author}{\bibfnamefont{A.}~\bibnamefont{Aviram}},
  \bibinfo{journal}{Nature} \textbf{\bibinfo{volume}{408}},
  \bibinfo{pages}{541} (\bibinfo{year}{2000}).

\bibitem[{\citenamefont{Koumura et~al.}(1999)\citenamefont{Koumura, Zijlstra,
  van Delden, Harada, and Feringa}}]{Koumura1999a}
\bibinfo{author}{\bibfnamefont{N.}~\bibnamefont{Koumura}},
  \bibinfo{author}{\bibfnamefont{R.}~\bibnamefont{Zijlstra}},
  \bibinfo{author}{\bibfnamefont{R.}~\bibnamefont{van Delden}},
  \bibinfo{author}{\bibfnamefont{N.}~\bibnamefont{Harada}}, \bibnamefont{and}
  \bibinfo{author}{\bibfnamefont{B.}~\bibnamefont{Feringa}},
  \bibinfo{journal}{Nature} \textbf{\bibinfo{volume}{401}},
  \bibinfo{pages}{152} (\bibinfo{year}{1999}).

\bibitem[{\citenamefont{Browne and Feringa}(2006)}]{Browne2006a}
\bibinfo{author}{\bibfnamefont{W.}~\bibnamefont{Browne}} \bibnamefont{and}
  \bibinfo{author}{\bibfnamefont{B.}~\bibnamefont{Feringa}},
  \bibinfo{journal}{Nat. Nanotech.} \textbf{\bibinfo{volume}{1}},
  \bibinfo{pages}{25} (\bibinfo{year}{2006}).

\bibitem[{\citenamefont{Park et~al.}(2000)\citenamefont{Park, Park, Lim,
  Anderson, Alivisatos, and McEuen}}]{Park2000a}
\bibinfo{author}{\bibfnamefont{H.}~\bibnamefont{Park}},
  \bibinfo{author}{\bibfnamefont{J.}~\bibnamefont{Park}},
  \bibinfo{author}{\bibfnamefont{A.~K.~L.} \bibnamefont{Lim}},
  \bibinfo{author}{\bibfnamefont{E.~H.} \bibnamefont{Anderson}},
  \bibinfo{author}{\bibfnamefont{A.~P.} \bibnamefont{Alivisatos}},
  \bibnamefont{and} \bibinfo{author}{\bibfnamefont{P.~L.}
  \bibnamefont{McEuen}}, \bibinfo{journal}{Nature}
  \textbf{\bibinfo{volume}{407}}, \bibinfo{pages}{57 } (\bibinfo{year}{2000}).

\bibitem[{\citenamefont{Beton et~al.}(1995)\citenamefont{Beton, Dunn, and
  Moriarty}}]{Beton1995a}
\bibinfo{author}{\bibfnamefont{P.~H.} \bibnamefont{Beton}},
  \bibinfo{author}{\bibfnamefont{A.~W.} \bibnamefont{Dunn}}, \bibnamefont{and}
  \bibinfo{author}{\bibfnamefont{P.}~\bibnamefont{Moriarty}},
  \bibinfo{journal}{Appl. Phys. Lett.} \textbf{\bibinfo{volume}{67}},
  \bibinfo{pages}{1075} (\bibinfo{year}{1995}).

\bibitem[{\citenamefont{Barth et~al.}(2005)\citenamefont{Barth, Costantini, and
  Kern}}]{Barth2005a}
\bibinfo{author}{\bibfnamefont{J.~V.} \bibnamefont{Barth}},
  \bibinfo{author}{\bibfnamefont{G.}~\bibnamefont{Costantini}},
  \bibnamefont{and} \bibinfo{author}{\bibfnamefont{K.}~\bibnamefont{Kern}},
  \bibinfo{journal}{Nature} \textbf{\bibinfo{volume}{437}},
  \bibinfo{pages}{671} (\bibinfo{year}{2005}).

\bibitem[{\citenamefont{Berninger et~al.}(2007)\citenamefont{Berninger,
  St\'{e}fanov, Deachapunya, and Arndt}}]{Berninger2007a}
\bibinfo{author}{\bibfnamefont{M.}~\bibnamefont{Berninger}},
  \bibinfo{author}{\bibfnamefont{A.}~\bibnamefont{St\'{e}fanov}},
  \bibinfo{author}{\bibfnamefont{S.}~\bibnamefont{Deachapunya}},
  \bibnamefont{and} \bibinfo{author}{\bibfnamefont{M.}~\bibnamefont{Arndt}},
  \bibinfo{journal}{Phys. Rev. A} \textbf{\bibinfo{volume}{76}},
  \bibinfo{pages}{013607} (\bibinfo{year}{2007}).

\bibitem[{\citenamefont{Patorski}(1989)}]{Patorski1989a}
\bibinfo{author}{\bibfnamefont{K.}~\bibnamefont{Patorski}}, in
  \emph{\bibinfo{booktitle}{Progress in Optics {XXVII}}}, edited by
  \bibinfo{editor}{\bibfnamefont{E.}~\bibnamefont{Wolf}}
  (\bibinfo{publisher}{Elsevier}, \bibinfo{address}{Amsterdam},
  \bibinfo{year}{1989}), pp. \bibinfo{pages}{2--108}.

\bibitem[{\citenamefont{Brezger et~al.}(2003)\citenamefont{Brezger, Arndt, and
  Zeilinger}}]{Brezger2003a}
\bibinfo{author}{\bibfnamefont{B.}~\bibnamefont{Brezger}},
  \bibinfo{author}{\bibfnamefont{M.}~\bibnamefont{Arndt}}, \bibnamefont{and}
  \bibinfo{author}{\bibfnamefont{A.}~\bibnamefont{Zeilinger}},
  \bibinfo{journal}{J. Opt. B: Quantum Semiclass. Opt.}
  \textbf{\bibinfo{volume}{5}}, \bibinfo{pages}{S82} (\bibinfo{year}{2003}).

\end{thebibliography}

\end{document}